\documentclass[conference]{IEEEtran}
\IEEEoverridecommandlockouts
\usepackage{cite}
\usepackage{amsmath,amssymb,amsfonts}
\usepackage{algorithmic}
\usepackage{graphicx}
\usepackage{textcomp}
\usepackage{xcolor}
\def\BibTeX{{\rm B\kern-.05em{\sc i\kern-.025em b}\kern-.08em
    T\kern-.1667em\lower.7ex\hbox{E}\kern-.125emX}}

\usepackage{amsmath} 
\usepackage{amsthm}
\usepackage{graphicx}
\usepackage{epstopdf}
\usepackage{alltt}
\usepackage{multirow}
\usepackage{makecell}
\usepackage{hhline}
\usepackage{lipsum}
\usepackage{algorithmic}
\usepackage{color}
\usepackage{makecell}
\usepackage{amssymb}
\usepackage{pifont}
\usepackage{stfloats}
\usepackage{multirow}
\newcommand{\cmark}{\ding{51}}%
\newcommand{\xmark}{\ding{55}}%
\DeclareGraphicsExtensions{.eps}
\usepackage[]{algorithm2e}
\usepackage{textcomp}
\usepackage{makecell}
\usepackage{mathtools}
\usepackage[T1]{fontenc}
\usepackage{upquote}
\usepackage[caption=false]{subfig}

\begin{document}

\title{Global Roaming Trust-based Model for V2X Communications\\
{}
\thanks{A. Alnasser (alalnasser@ksu.edu.sa) is also with School of Engineering and Computing Sciences,  Durham University, Durham, UK. }
}

\author{\IEEEauthorblockN{Aljawharah Alnasser}
\IEEEauthorblockA{\textit{Department
		of Information Technology} \\
\textit{King Saud University}, Riyadh, Saudi Arabia}
\and
\IEEEauthorblockN{Hongjian Sun}
\IEEEauthorblockA{\textit{Department of Engineering} \\
\textit{Durham University}, Durham, UK}
}

\maketitle

\begin{abstract}
Smart cities need to connect physical devices as a network to improve the efficiency of city operations and services. Intelligent Transportation System (ITS) is one of the key components in smart cities, due to its capability of supporting communications between vehicles to improve the driving experience.
Whilst Vehicle-to-Everything (V2X) communications are essential, cyber-security poses a significant challenge in V2X communications. A V2X communication link is vulnerable to various cyber-attacks including internal and external attacks. Internal attacks cannot be detected by conventional security schemes because the compromised nodes have valid credentials. Thus, a new trust model is urgently needed to mitigate cyber-security risks. In this paper, a global roaming trust-based security model is proposed for V2X communications.  Each vehicle has a global knowledge about malicious nodes in the network. In addition, various experiments are conducted with different percentage of malicious nodes to measure the performance of the proposed model.  Simulation results show that the proposed model improves False Negative Rate (FNR) by 33.5\% in comparison with the existing method.
\end{abstract}


\section{Introduction}
Intelligent Transportation System (ITS) is one of the leading smart systems which have been developed to obtain reliable transportation. One vehicle can establish a communication with other vehicles and/or infrastructure units using Vehicle-to-Everything (V2X) communications. Vehicles include all moving road entities, such as cars, bicycles, buses, trains and motorcycles. The road entity periodically broadcasts a message which contains status information, such as speed, directions and location. V2X supports several types of communication links as shown in Fig.1, e.g. Vehicle-to-Vehicle (V2V), Vehicle-to-Pedestrian (V2P), Vehicle-to-Grid (V2G) and Vehicle-to-Infrastructure (V2I).

As a consequence, the communication link between road entities is exposed to either internal or external cyber-attacks. \textit{External attacks} means that unauthorized nodes launch the malicious behavior. Fortunately, the network can be protected against these attacks by applying conventional security schemes, such as encryption and authentication. \textit{Internal attacks} means that authorized nodes initiate the malicious behavior. Unfortunately, the compromised nodes are hard to be detected because they have valid credentials. As a result, a trust-based model was studied to protect the network against internal attacks in \cite{Trust1}, by continuously monitoring the surrounding nodes\textquotesingle\ behavior. When a misbehavior node is detected, a warning alarm is sent to the network \cite{general2}. 

\begin{figure}[t!]
	\centering
	\includegraphics[width=\linewidth]{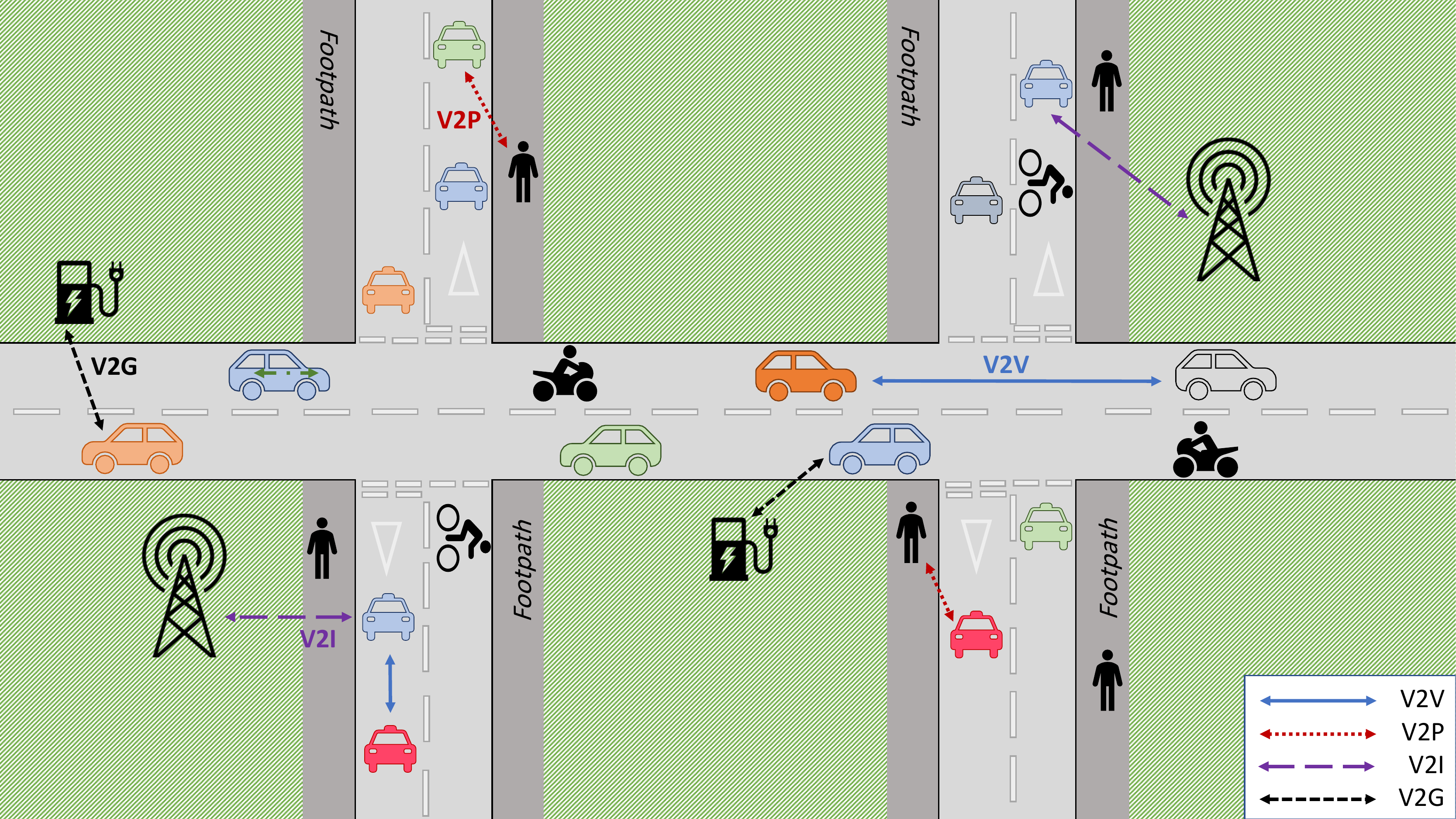}
	\caption{V2X communication links}
\end{figure}

There is a rich literature on developing security models to provide data confidentiality in V2X communications. For instance, Liu \textit{et al.} \cite{r3} designed a privacy-preserving ad conversion protocol for V2X-assisted proximity marketing that achieves input certification and output verifiability against malicious ad networks.  Ulybyshev \textit{et al.} \cite{r4} suggested a data exchange method for V2X communications, to ensure data confidentiality and integrity. This method supports encrypted search over encrypted vehicle records that could be stored in untrusted cloud. Simplicio \textit{et al.} \cite{r5} improved the structure of SCMS\textquotesingle s certificate revocation and linkage approach by addressing some limitations. The proposed modifications support the temporary revocation and linkage of pseudonym certificates. Furthermore, Cheng \textit{et al.} \cite{r7} presented a remote attestation security model based on a privacy-preserving blockchain. The model is comprised of two parts:  identity authentication and the calculation of the nodes to make final decisions and write them into data blocks. 

Recently, the authentication of V2X communications has been well studied. For instance, Yang \textit{et al.} \cite{vx1} implemented an authentication model for V2X communications. This model consists of two schemes: one scheme for V2V communications, and another for V2I communications. Villarreal-Vasquez \textit{et al.} \cite{vx3} proposed a dynamic approach which achieves the trade-off between safety, security and performance of V2X systems. However, the analysis is limited to V2V communications compliant with IEEE802.11p. 
In addition, Kiening \textit{et al.} \cite{vx4} studied the security requirements for V2X systems in particular trust assurance levels. A certification framework was designed to support trust establishment between road entities in V2X communications. Indeed, the node should be trusted if it has been correctly authenticated. Ahmed and Lee \cite{general5} evaluated security services of the new LTE-based V2X architecture. Building on evaluation results, a practical solution was proposed to protect privacy and achieve security requirements of message exchange in V2X networks. Also, Jung \textit{et al.}\cite{r1} suggested a procedure and test scenario to achieve secure communication for autonomous cooperation driving. 
Furthermore, there are some research on ensuring data integrity. To defend against both false data injection and packet drop attacks, a new model was proposed in \cite{r2} that particularly focuses on the security in sensing systems for V2X networks. However, far less effort has been devoted to defending against internal attacks. 

To deal with internal attacks, this paper studies a global roaming trust-based model for V2X communications.  The performance of the proposed model is then evaluated by comparing it with an existing model \cite{J3}. The simulation results show that the proposed model outperforms the existing one. This paper makes two main contributions to the field of vehicular network security:
\begin{itemize}
	\item This paper proposes a global roaming trust-based model for V2X communications. Different from existing research, the nodes have global knowledge about malicious nodes in the network. 	
	\item This paper compares the performance of the proposed model with the existing model in \cite{J3}; the proposed model improves the False Negative Rate (FNR) by 33.5\% when the percentage of malicious nodes is around 87.5\%.

\end{itemize}

The remaining of this paper is organised as follows. Section II presents the system model. Section III provides a detailed description of the proposed trust model. Section IV includes both simulation setup and experimental results. Section V focuses on performance comparison with the existing model \cite{J3}. Section VI draws conclusions.

	\section{System Model}

The considered network consists of $N$ road entities, which move at various speeds, and $M$ fixed Road Side Units (RSUs). Each road entity sends three types of messages: \textit{Beacon message} which is sent periodically to inform the surrounding nodes about its current speed, location and direction; \textit{transaction message} which contains confidential information and it is sent to the core network; and \textit{warning message} that is sent to the surrounding RSUs when a malicious node is detected. Each time the road entity sends a message to the core network, it should go through the following phases:
\begin{itemize}
	\item \textit{Connectivity phase:} each road entity examines its connectivity with the core network and the surrounding entities.
	\item \textit{Communication phase:} if the source entity has a connection with the core network, it forwards its packet to the nearest RSU. Otherwise, the packet is sent to a trusted entity to relay them to the core network.
\end{itemize}
Moreover, the considered network has two types of nodes which are normal and malicious nodes. The normal node keeps monitoring the surrounding environment and sends its packets to the core network. Also, it relays any received packet to the nearest RSU. On the other hand, the malicious node launches various attacks to disturb the network performance such as:

%



\begin{itemize}
	\item Selective forwarding attack: occurs when the malicious node drops some of the received packets randomly to escape punishment. 
	\item Recommendation attack: occurs when the malicious node sends bogus recommendations regarding other nodes:
	\begin{itemize}
		\item In good-mouthing attack, the malicious node $f$ sends good recommendations regarding other malicious nodes $h_1, h_2, ...$ $h_{np}$ as shown in Fig.2(a). In this attack, the malicious nodes $h$ could be considered as normal nodes. Thus, the malicious node $f$ disturbs the decision phase.
		\item In bad-mouthing attack, the malicious node $f$ sends bad recommendations regarding other normal nodes $q_1, q_2, ...$ $q_{np}$ as shown in Fig.2(b). In this attack, the normal nodes $q$ may be classified by node $i$ as malicious nodes. 
	\end{itemize}
\end{itemize}
\begin{figure}[t!]
	\centering
	\includegraphics[width=\linewidth]{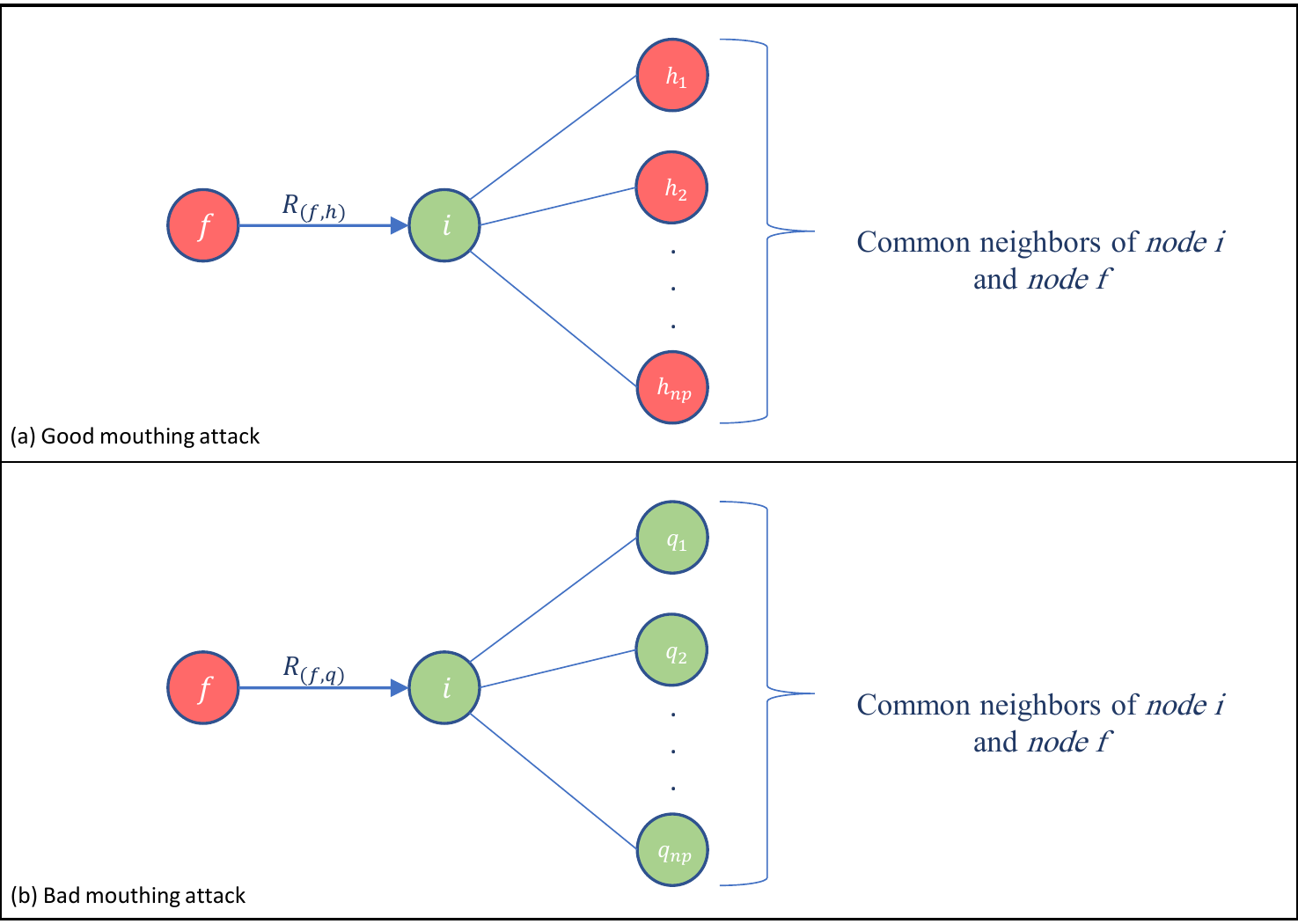}
	\caption{General model for recommendation attacks}
	
\end{figure}

\section{Global Roaming Trust-based Model}

The global roaming trust-based model maintains two levels of trust as shown in Fig.3: \textit{road entities level} and \textit{RSU level}. The road entity evaluates the trustworthiness of surrounding entities, and then sends warning messages to the surrounding RSUs when a malicious node is detected. When the RSUs receive high volume of warning messages from the surrounding entities, they generate an alarm and send it to the central unit. The details of this model are presented as follows.

\begin{table*}[bp]
	\caption{Local trust computation of the proposed model}
	\begin{center}
		\begin{tabular}{|c|c|c|c|c|c|c|}
			\hline
			& \makecell{Existing of current communication \\between node i and node j} &\makecell{Existence of the recommendations \\about node j}&$w_1$&$Trust_1$&$w_2$&$Trust_2$\\
			\hline
			\multirow{4}{*}{\rotatebox{90}{\makecell{First time \\  communication}}}&\rule{0pt}{12pt}\cmark&\cmark&Eq.(7)&$T_{indirect(i,j)}^{(t)}$&$1-w_1$&$T_{direct(i,j)}^{(t)}$\\
			\cline{2-7}
			&\rule{0pt}{12pt}\cmark&\xmark&1&$T_{direct(i,j)}^{(t)}$&0&--\\
			\cline{2-7}
			&\rule{0pt}{12pt}\xmark&\cmark&1&$T_{indirect(i,j)}^{(t)}$&0&--\\
			\cline{2-7}
			&\rule{0pt}{12pt}\xmark&\xmark&1&$T_{l(i,j)}^{(0)}$&0&--\\
			\hline
			\hline
			\multirow{4}{*}{\rotatebox{90}{\makecell{Have Previous \\Communication}}}&\rule{0pt}{12pt}\cmark&\cmark&Eq.(7)&$T_{indirect(i,j)}^{(t)}$&$1-w_1$&$T_{current(i,j)}^{(t)}$\\
			\cline{2-7}
			&\rule{0pt}{12pt}\cmark&\xmark&1&$T_{current(i,j)}^{(t)}$&0&-\\
			\cline{2-7}
			&\rule{0pt}{12pt}\xmark&\cmark&Eq.(7)&$T_{indirect(i,j)}^{(t)}$&$1-w_1$&$T_{past(i,j)}^{(t)}$\\
			\cline{2-7}
			&\rule{0pt}{12pt}\xmark&\xmark&1&$T_{past(i,j)}^{(t)}$&0&--\\
			\hline
		\end{tabular}
	\end{center}
	\label{tab:multicol}
\end{table*}

\subsection {Road entity level}
During time interval $t$, each road entity measures the trustworthiness of all surrounding entities. Indeed, node $i$ continuously monitors its one-hop neighbors $j$. Then, node $i$ is able to compute direct trust using the collected information. In addition, node $i$ sends recommendation requests to the surrounding nodes $k$ regarding node $j$.  The proposed model manages two trust components as follows.
\begin{itemize}
	\item \textbf{Current Trust - $T_{current(i,j)}^{(t)}$}: it is computed by node $i$ to evaluate the communication experience with node $j$ during time interval $t$. It is calculated using
	\begin{equation}
	T_{current(i,j)}^{(t)} = \frac{T_{past(i,j)}^{(t)}+T_{direct(i,j)}^{(t)}}{2}
	\end{equation}
	It is measured based on the following trust values:
	\begin{itemize}
		\item \textit{Past trust - $T_{past(i,j)}^{(t)}$}: it is a measure for the past behavior of node $j$. The past trust is considered to prevent the non-continuous malicious behavior.
		
		\item \textit{Direct trust - $T_{direct(i,j)}^{(t)}$}: it is an evaluation for the communication experience with the neighboring nodes $j$. It is computed using 
		\begin{equation}
		T_{direct(i,j)}^{(t)} = \frac{Successful\_Interactions}{Total\_Interactions}
		\end{equation}
		where $Successful\_Interactions$ is the number of successful interactions between node $i$ and node $j$, and $Total\_Interactions$ is the total number of interactions between node $i$ and node $j$.
	\end{itemize}
	\item \textbf{Indirect Trust - $T_{indirect(i,j)}^{(t)}$}: it is a measure for the behavior of neighboring nodes $j$ using surrounding nodes\textquotesingle\ opinions. Node $i$ collects recommendations from the surrounding nodes regarding node $j$. Before computing indirect trust, node $i$ applies the following steps:
	\begin{itemize}
		\item \textit{Confidence value computation- $C_{(i,k)}^{(t)}$}: node $i$ measures the confidence value for each recommender node $k$.    $C_{(i,k)}^{(t)}$ is computed by
			\begin{figure}[t!]
			\centering
			\includegraphics[width=\linewidth]{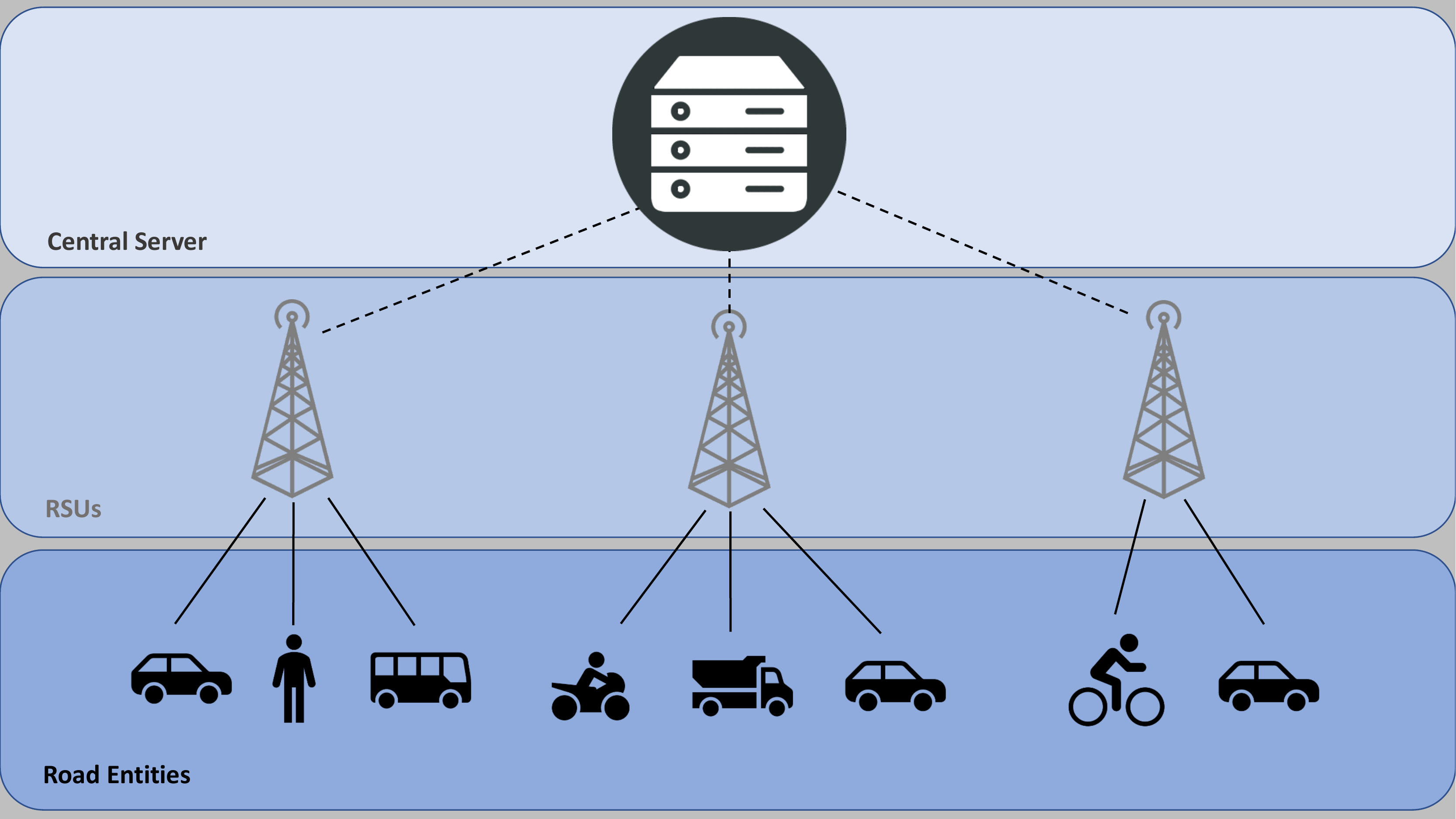}
			\caption{Trust levels in the proposed model}
			
		\end{figure}
		\begin{equation} 
		C_{(i,k)}^{(t)}=\begin{cases}
		1, & \text{if $T_{l(i,k)}^{(t)}\geq Th_{max}$}.\\
		C_{w}, & \text{if $Th_{min} \leq T_{l(i,k)}^{(t)} < Th_{max}$}. \\
		0, & \text{if $T_{l(i,k)}^{(t)} < Th_{min}$}.
		\end{cases}
		\end{equation}
		where $C_{w}$ is the confidence weight for uncertain recommendations. 
		\item \textit{Recommendations clustering}: node $i$ classifies the received recommendations into 
		two groups which are positive and negative recommendations using $Th_{min}$. 
	\end{itemize}
 
 After that, each node $i$ calculates indirect trust for node $j$ by applying different weights $\alpha$ and $\beta$ for $P_{(i,j)}^{(t)}$ and $N_{(i,j)}^{(t)}$ respectively. It is calculated using
%
\begin{equation}
T_{indirect(i,j)}^{(t)} = \alpha \times P_{(i,j)}^{(t)}+\beta \times N_{(i,j)}^{(t)}
\end{equation}
where $P_{(i,j)}^{(t)}$ is the average value of positive recommendations; and $N_{(i,j)}^{(t)}$ the average value of negative recommendations. The weights are computed by

\begin{equation}
\alpha=\frac{n}{n+m} , \beta=\frac{m}{n+m}
\end{equation}
where $n$ and $m$ are the number of positive and negative recommendations respectively.
	\item \textbf{Local Trust - $T_{l(i,j)}^{(t)}$}: each node $i$ is able to compute local trust for node $j$ and make a decision. Generally, local trust is computed using 
	\begin{equation}
	T_{l(i,j)}^{(t)} = w_1 \times Trust_1 + w_2 \times Trust_2
	\end{equation}
	
	where $Trust_1$ and $Trust_2$ are adjusted based on three factors which are the occurrence of current communications between node $i$ and node $j$; the existence of the recommendations about node $j$; and the presence of a previous connection between node $i$ and node $j$. The measurement of  $Trust_1$ and $Trust_2$ are described in Table I.
	
	In addition, trust weights $w_1$ and $w_2$ are changed based on recommendation factor ($RC$) and the number of neighbors. $w_1$ and $w_2$ are weights for indirect trust and (direct/current or past) trust respectively. $w_1$ represents the recommendation rate as follows:

	\begin{equation}
	w_1 = (m+n) \times \frac{RC}{Neighbors^{(t)}} 	
	\end{equation}
	where $w_2 = 1-w_1$, and $Neighbors^{(t)}$ is the number of node $i$ neighbors at time $t$. 

	\item \textbf{Local decision}: node $i$ has a local blacklist which has a list of malicious nodes based on the local decision. Thus, node $i$ stops the communication with any node $j$ in the blacklist. The decision is made using

	\begin{equation} 
	D_{Local}=\begin{cases}
	Trusted, & \text{if $T_{l(i,j)}^{(t)}\geq Th_{max}$}.\\
	Uncertain, & \text{if $Th_{min} \leq T_{l(i,j)}^{(t)}< Th_{max}$}.\\
	Malicious, & \text{if $T_{l(i,j)}^{(t)} < Th_{min}$}.
	\end{cases}
	\end{equation}
	where $Th_{min}$ and $Th_{max}$ are minimum and maximum trust thresholds, respectively.  After that, the node updates its local blacklist and sends malicious and uncertain warning messages to the surrounding RSUs.
\end{itemize}

\subsection {RSU level}
During time interval $t^{\prime}$, where $t^{\prime}>t$, RSUs start trust calculation phase. First, each RSU measures the percentage of malicious and uncertain alarms regarding node $j$ using
\begin{equation}
M= \frac{m^{\prime}}{t^{\prime}} , \ \ \ \ \   U= \frac{u}{t^{\prime}}
\end{equation}
where $m^{\prime}$ and $u$ are the number of malicious and uncertain warnings respectively. Second, each RSU is able to make a decision regarding node j using 
\begin{equation}
Decision_{j}= Rate_{M} - Rate_{U} 
\end{equation}
where $Rate_{M}$ and $Rate_{U}$ are the rates of malicious alarms and uncertain alarms respectively. They are calculated using 
\begin{equation}
Rate_{M}= \frac{M}{M+U} 
\end{equation}

\begin{equation}
Rate_{U}= \frac{U}{M+U}
\end{equation}

Finally, the RSU classifies node $j$ as malicious node when $Decision_{j}>0$. Therefore, RSU sends malicious alarm to the central server. 

\subsection {Global Trust decision}
At this stage, central server can make global decision regarding node $j$ based on the alarms which are received from RSUs. 
\begin{equation} 
D_{Global}=\begin{cases}
Malicious, & \text{if $A_{m} \geq \frac{Total\_RSUs}{2}-1$}.\\
Normal, & \text{Otherwise}.
\end{cases}
\end{equation}
where $A_{m}$ is the number of malicious warnings that are received regarding node $j$. Node $j$ is added to the global blacklist when it is classified as malicious node. Central server broadcasts the updated global blacklist to RSUs. Then, RSUs rebroadcast it again to all roads entities that are covered by the network. The road entities updates the local blacklist based on the received global blacklist.

\section{Simulation Analysis}
This section describes the simulation setup for evaluating the performance of the proposed model. The effect of changing parameters on the false alarm rate is also analysed.  

\subsection{Network specifications}
We used MATLAB R2016b to conduct the simulation of a V2X network with 24 road entities and 9 RSUs with parameters as shown in Table II. The road entities move over an area of $900 \times 900$ $m^2$ with various speed ranges. The considered area is composed of two intersections using three two-lanes roads. The road entity sends the transaction message to the core network directly or using a multi-hop routing protocol. 
To measure the performance of the proposed trust model, we study various types of malicious nodes: six selective forwarding attackers, three good-mouthing attackers and three bad-mouthing attackers. 


\begin{table}[t!]
	\centering
	\caption {Simulation Parameters} \label{tab:title} 
	\begin{center}
		
		\begin{tabular}{ |c|c| } 
			\hline
			Parameter & Value  \\ 
			\hline
			Simulation time (T) & 100 iteration \\ 
			\hline
			Speed ranges & \makecell{Vehicle:(10-30) m/s,\\ Pedestrians:(0-8) m/s \\ Cycles:(3-10) m/s,\\ Motorcycle:(10-30) m/s} \\ 
			\hline
			Number of nodes & 24 nodes \\ 
			\hline
			$Th_{max}$ & 0.7 \\ 
			\hline
			$Th_{min}$ & 0.4 \\ 
			\hline
			$RC$ & 0.3 \\
			\hline
			$C_{w}$ & 0.9 \\
			\hline
			$T_{l(i,j)}^{(0)}$ & 0.5 \\
			\hline
		\end{tabular}
	\end{center}
\end{table}


\subsection{Simulation Results}
In this section, we study the impact of changing parameters on the global trust measure and relate these to the false alarm rate. False alarm rate includes False Negative Rate (FNR) and False Positive Rate (FPR). FNR measures the rate of undetected attacks, whilst FPR measures the rate of classifying normal nodes as malicious. We run the simulations using the initial parameters $Th_{max}=0.9, RC=0.3, C_{w}=0.9$. Then, we updated their values with the optimal ones. 
%
%

\begin{figure}[b!]
	\centering
	\subfloat[Minimum threshold ($Th_{min}$)]{%
		\includegraphics[width=0.8\columnwidth]{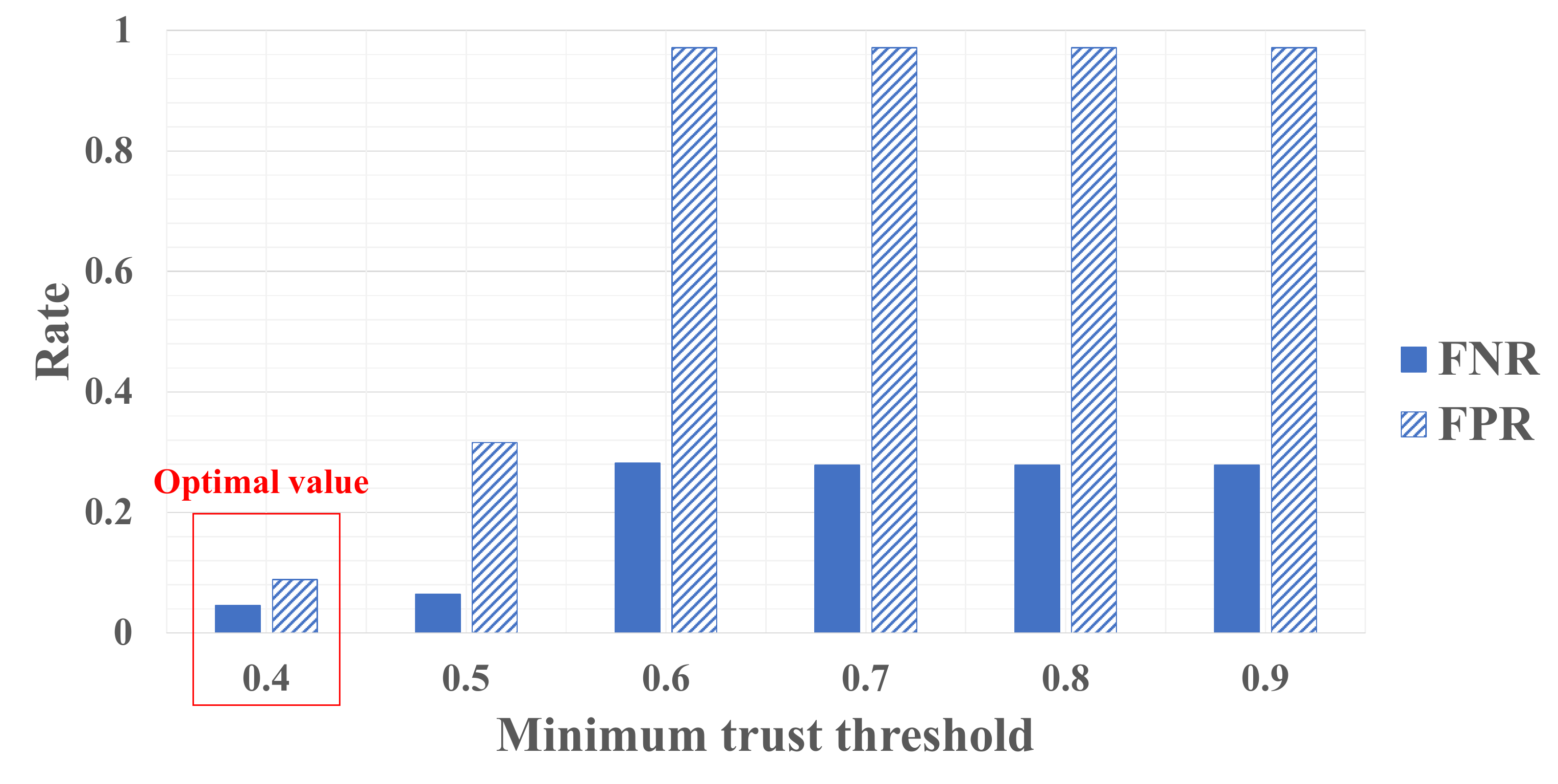}%
	}
	
	\subfloat[Maximum threshold ($Th_{max}$)]{%
		\includegraphics[width=0.8\columnwidth]{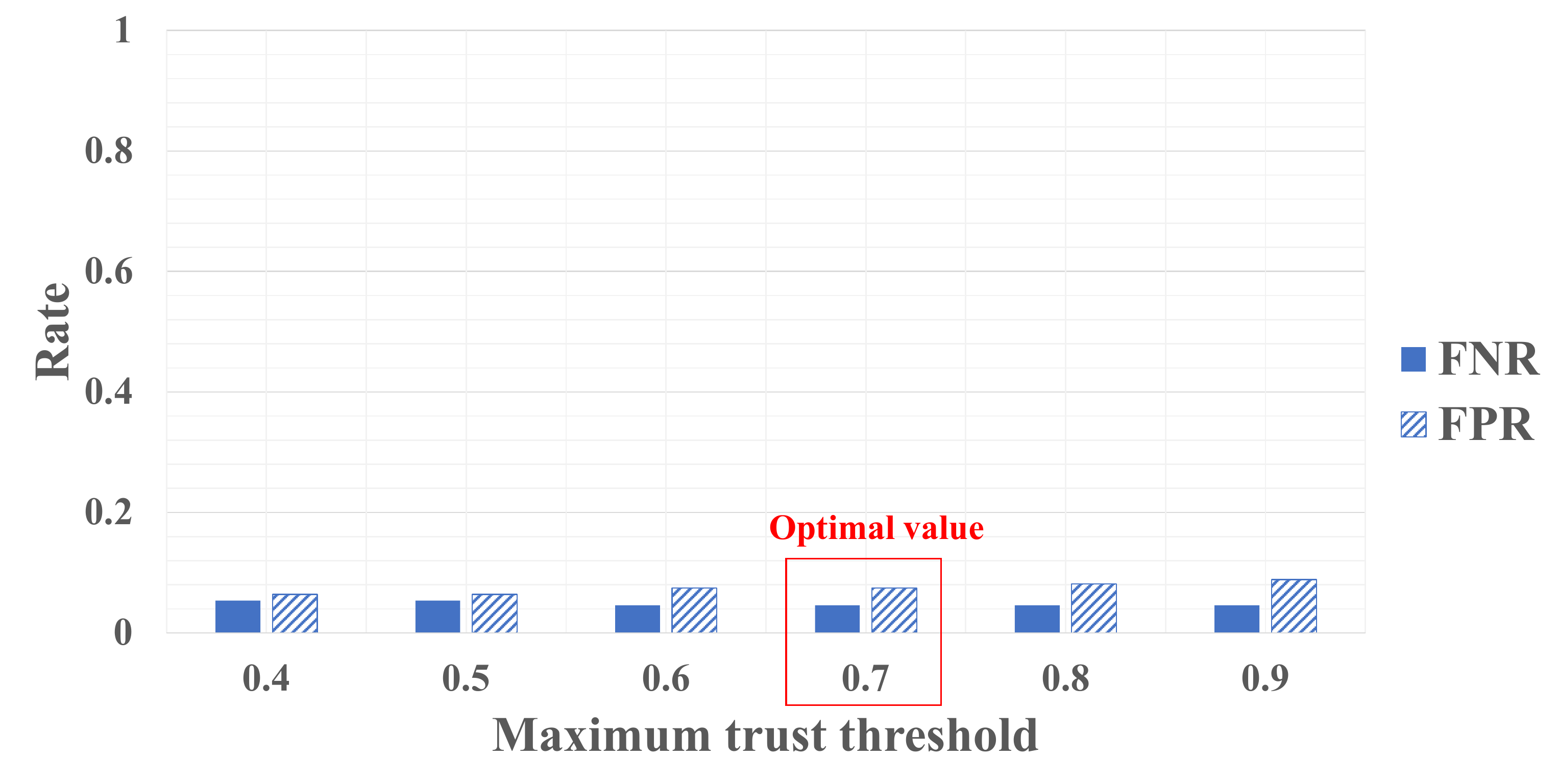}%
	}
	\caption{Effect of changing trust thresholds on false alarm rate}
	
\end{figure}
\subsubsection{Effect of trust thresholds on false alarm rate}  
The simulation experiments were run with initial parameters. We study how various values of $Th_{min}$ has an impact on false alarm rate. Also, it helps us to define the optimal value for $Th_{min}$. The corresponding results are shown in Fig.4 (a). The following remarks can be made:
\begin{itemize}
	\item FNR increases when the value of $Th_{min}$ increases;
	\item FPR rises significantly as long as the $Th_{min}$ increases;
	\item the impact of $Th_{min}$ is high on FPR because as long as $Th_{min}$ goes up that means the malicious range is expanded. As a result, many normal nodes are classified as malicious nodes;
	\item when $Th_{min}=0.4$, it achieves low FNR and FPR values. 
\end{itemize}

Moreover, we study how various values of $Th_{max}$ has an impact on false alarm rate. The experiment was run with initial parameters and $Th_{min}=0.4$. The corresponding results are shown in Fig.4 (b). We notice that FNR slightly decreases when the value of $Th_{max}$ increases, however, the FPR slightly goes up as long as the $Th_{max}$ increases. We update initial value of $Th_{max}$ with 0.7 which is the optimal value. 

\subsubsection{Effect of recommendation factor ($RC$)}  			
The simulation experiments were run with updated initial parameters. Here, we study the effect of various values of $RC$ on the false alarm rate. By inspecting Fig.5 (a), the following remarks can be made:
\begin{itemize}
	\item FPR goes up when the value of $RC$ increases to reach approximately 0.27, however, the FNR is stable while RC increases;
	\item the $RC$ has an impact on FPR only because $RC$ is a part of the calculation of indirect trust weight $w_1$. Therefore, giving high weight to indirect trust results high FPR. As a result, the model starts making false decisions regarding the normal nodes.
	\item we choose $RC=0.3$ as an optimal value which is the same as initial value. 
\end{itemize}

%
%
%
%

\begin{figure}[b!]
	\centering
	\subfloat[Recommendation factor ($RC$) in eq.(7)]{%
		\includegraphics[width=0.8\columnwidth]{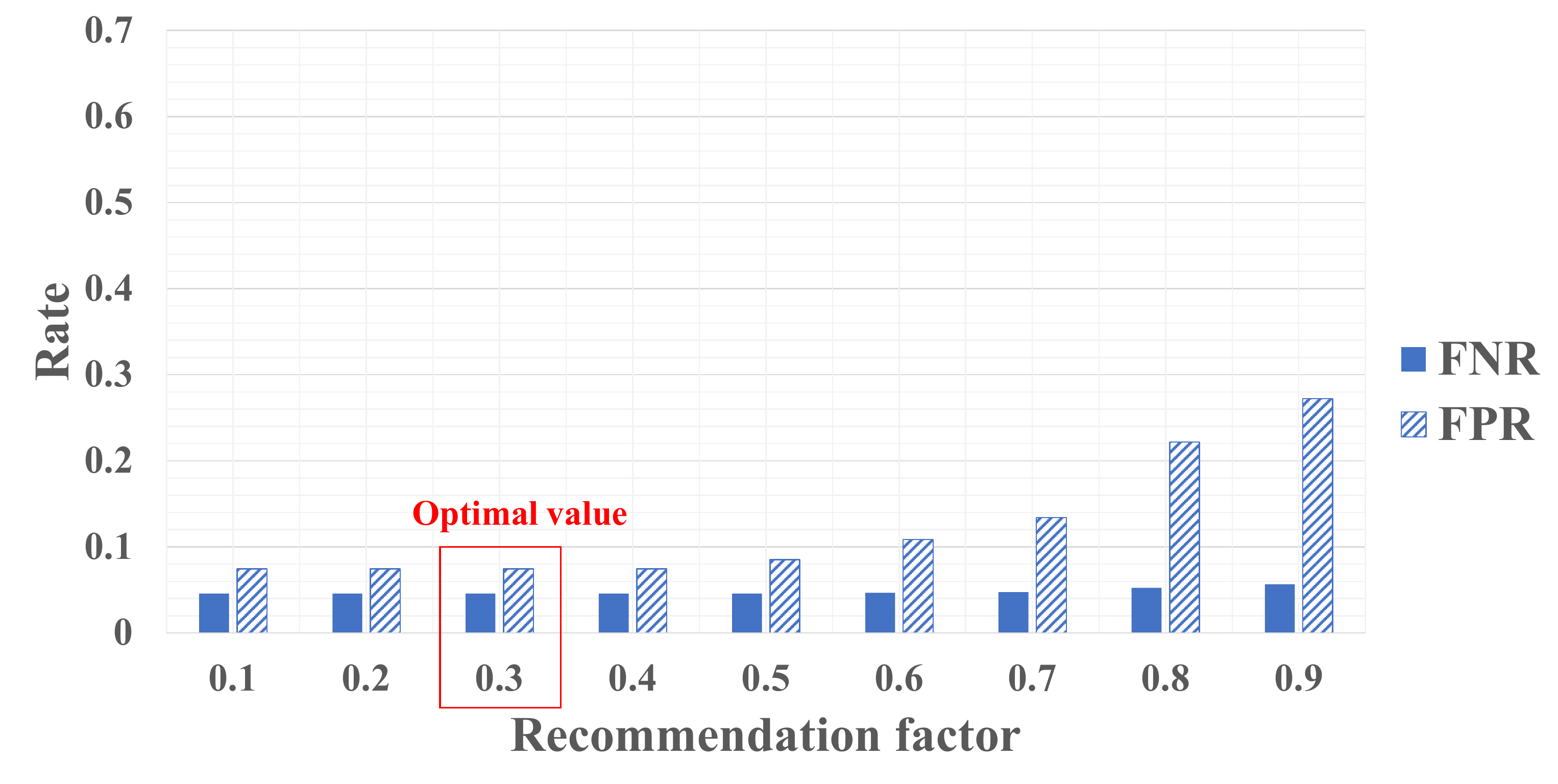}%
	}
	
	\subfloat[Confidence weight ($C_{w}$) in eq.(3)]{%
		\includegraphics[width=0.8\columnwidth]{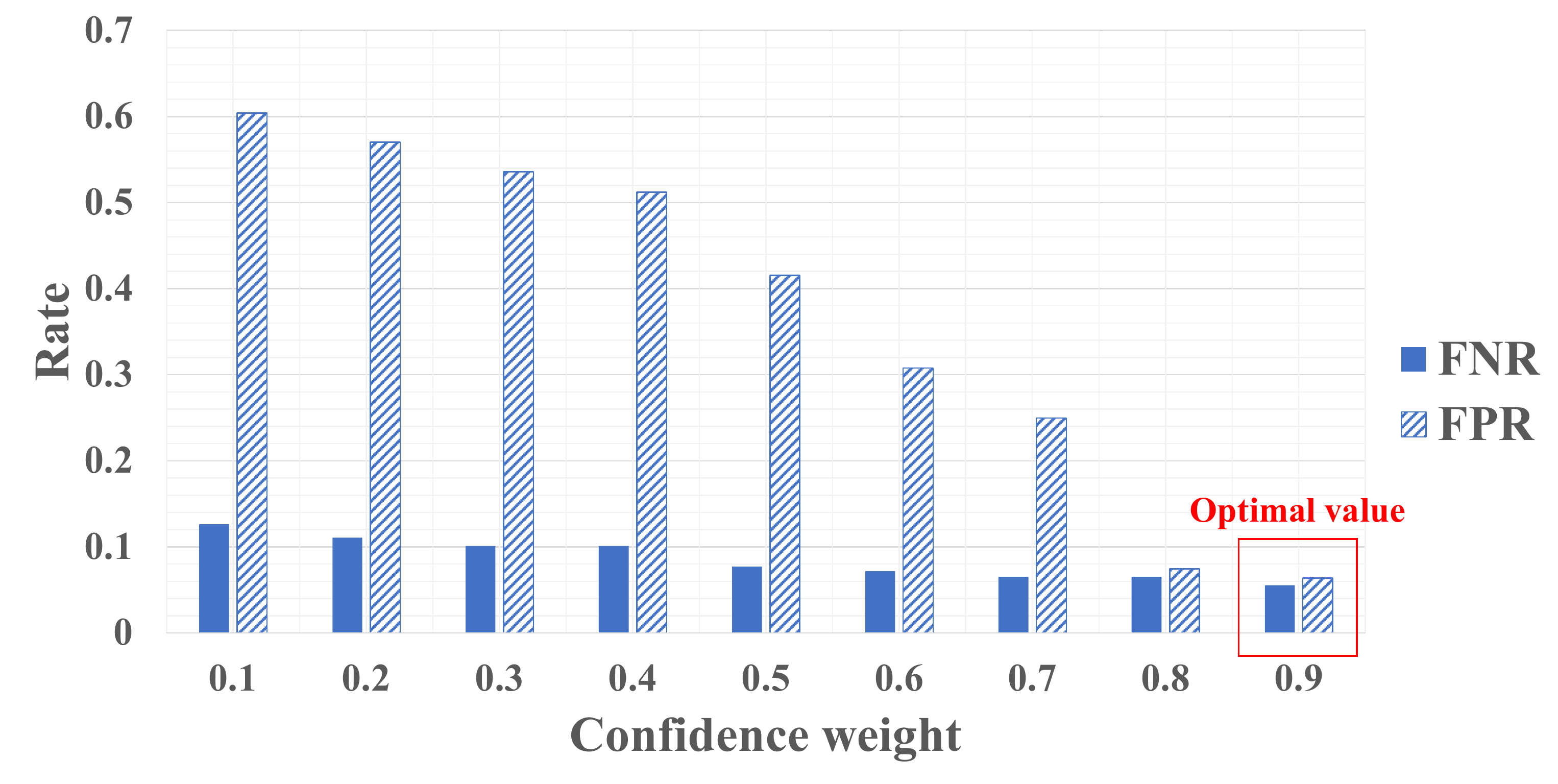}%
	}
	\caption{Effect of changing Recommendation factor and confidence weight on false alarm rate}
	
\end{figure}
\subsubsection{Effect of Confidence weight ($C_{w}$)}  			
We examine various values of $C_{w}$ to choose the value that achieves minimum false alarm rate, as shown in Fig.5 (b). Key findings are:
\begin{itemize}
	\item FPR goes down when the $C_{w}$ increases because we give lower weight for the recommendations that are sent by uncertain nodes, however, the FNR decreases slightly when the $C_{w}$ increases.
	\item majority of normal nodes are classified as uncertain, giving recommendations low weight results high FPR. 
	\item the initial value of $C_{w}$ is the optimal one.
\end{itemize}

%
\section{Performance Evaluation}
We use the existing model in \cite{J3} as a benchmark to evaluate the performance of the proposed model. The impact of various rates of malicious nodes on the false alarm rate is studied on the proposed model and existing model.

\subsection{Effect of selective forwarding attack on FNR}
Generally, when the model has a low FNR, it is able to detect the most malicious nodes. The result that is shown in Fig.6 (a) represents the FNR for various percentages of malicious nodes. The following remarks can be made:

\begin{itemize}
	\item in the existing model, the FNR reaches to 0.73 when the percentage of malicious nodes is equal to 87.50\%.
	\item FNR values in the proposed model is reduced. Thus, the global decision has the minimum FNR value for all rates of malicious nodes. 
\end{itemize}

%
%
%
%
%
%
\begin{figure}[b!]
	\centering
	\subfloat[FNR]{%
		\includegraphics[width=0.8\columnwidth]{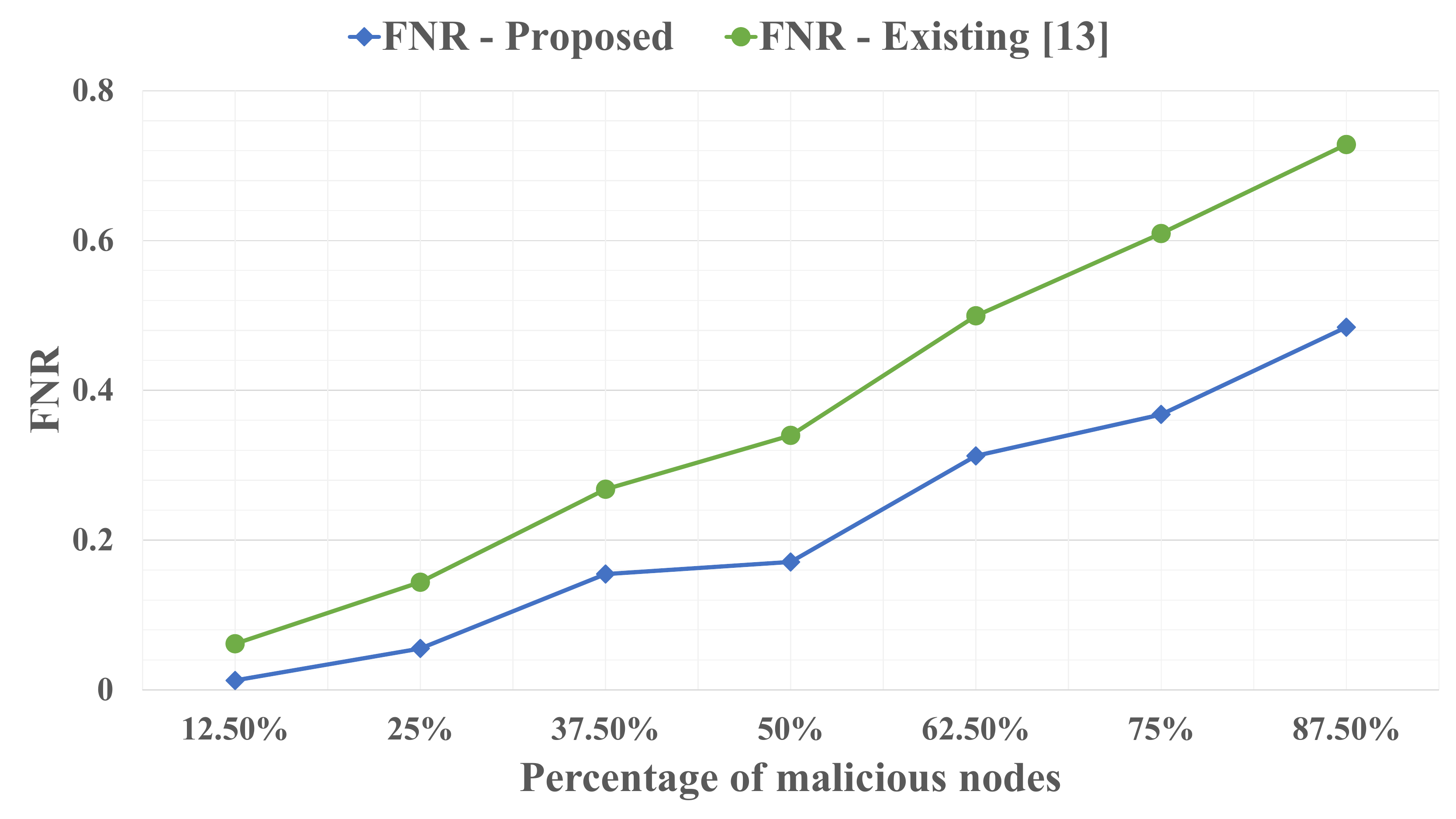}%
	}
	
	\subfloat[PDR]{%
		\includegraphics[width=0.8\columnwidth]{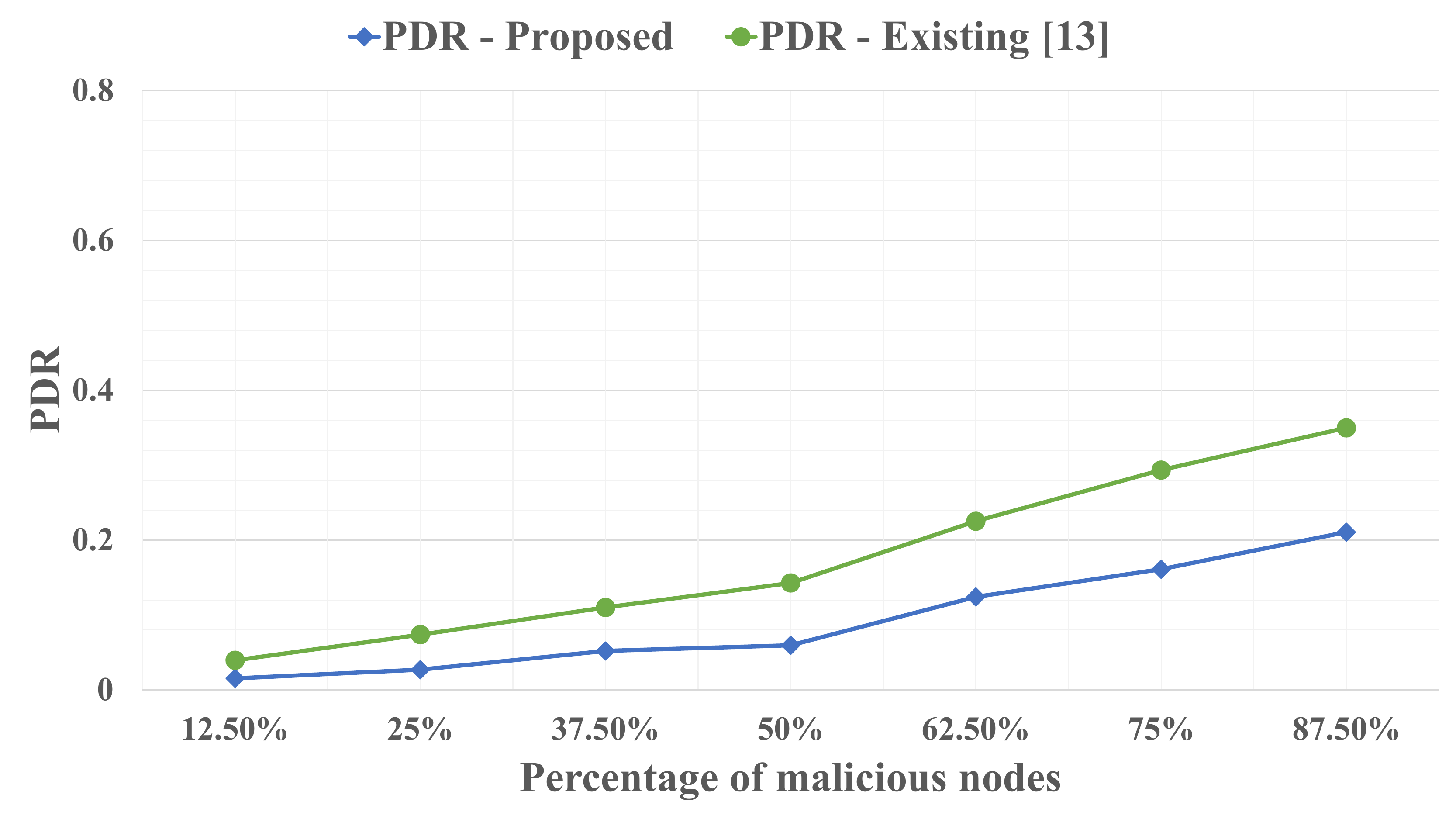}%
	}
	\caption{Effect of selective forwarding attack on FNR and PDR}
	
\end{figure}

%
%
%
%
\subsection{Effect of selective forwarding attack on PDR}
To measure the model performance, we measure the PDR with different percentage of malicious nodes as shown in Fig.6 (b). Generally, the PDR is increasing when the percentage of malicious nodes is increasing. In addition, the existing model produces high PDR which results from the high FNR. 
On the other hand, the proposed model has lower PDR which improves the network performance.
%
%
%
%
%
%
\subsection{Measuring the improvement rate}
We measure the improvement rate on FNR and PDR for the proposed model in comparison with the existing model \cite{J3} as shown in Fig.7. We notice that the FNR is highly improved in the proposed model when the percentage of malicious nodes is equal to 12.50\%. In addition, the rate at 50\%, which is a high percentage, increases again to around 50\%.

Moreover, we notice that the proposed model provides high improvement on PDR in comparison with the existing model, thus, it gains better network performance.

%


\begin{figure}[b!]
	\centering
	\includegraphics[width=\linewidth]{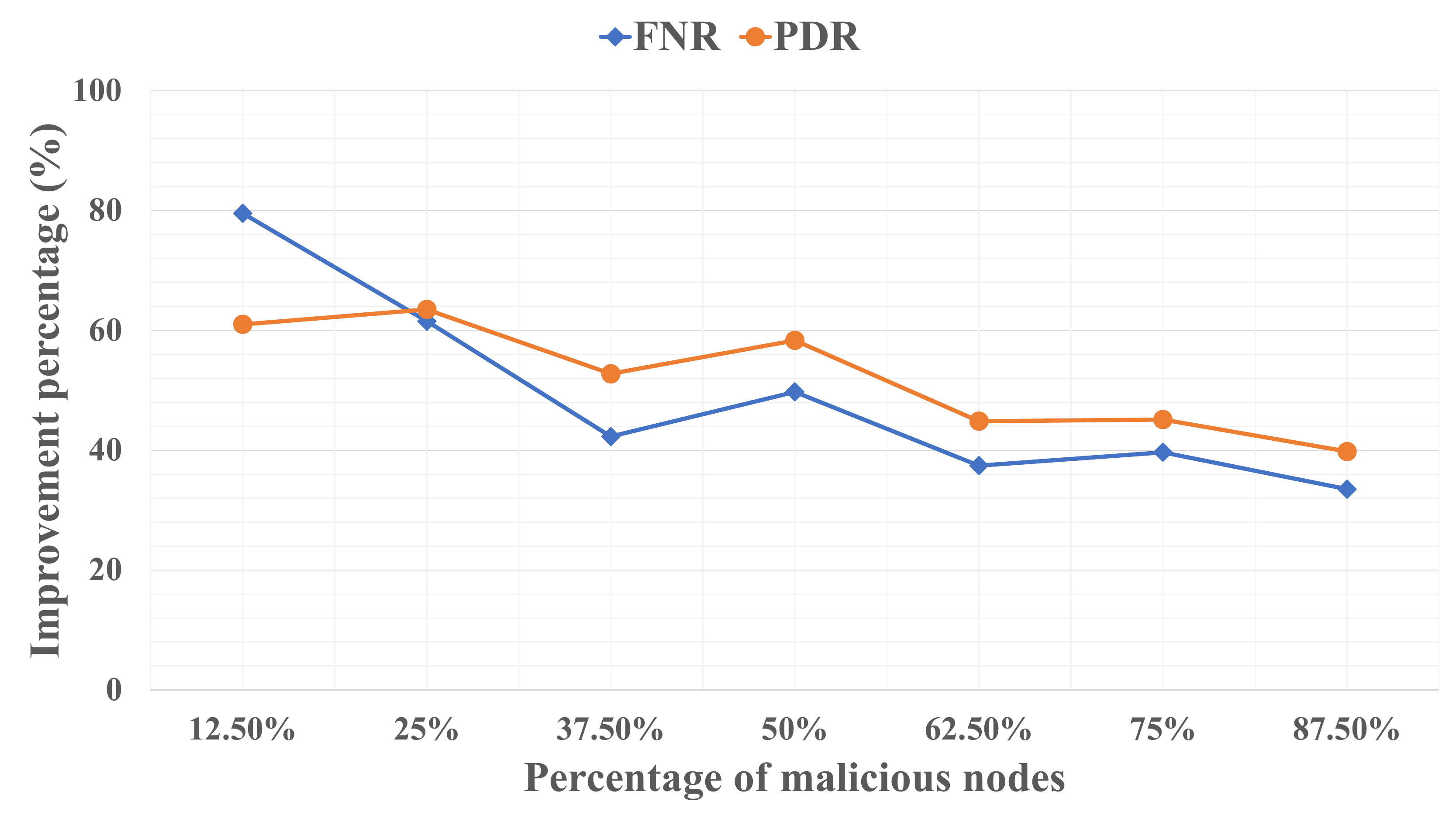}
	\caption{Improvement rate on FPR and PDR in the proposed model}
	
\end{figure}

\section{Conclusion}
In this paper, we proposed a global roaming trust-based model for the V2X network. Various malicious behaviors are considered to study the performance of the proposed model which are selective forwarding attack, bad-mouthing attack and good-mouthing attack. We conducted various experiments with different percentage of malicious nodes. Comparison results showed that the proposed model improved FNR by 33.5\% and PDR by 40\% when the percentage of malicious nodes is equal to 87.50\%.
In future work, we will improve the proposed model to consider RSU attacks.


\bibliographystyle{IEEEtran}
\bibliography{IEEEabrv,V2X-Ref}

\begin{thebibliography}{10}
\providecommand{\url}[1]{#1}
\csname url@samestyle\endcsname
\providecommand{\newblock}{\relax}
\providecommand{\bibinfo}[2]{#2}
\providecommand{\BIBentrySTDinterwordspacing}{\spaceskip=0pt\relax}
\providecommand{\BIBentryALTinterwordstretchfactor}{4}
\providecommand{\BIBentryALTinterwordspacing}{\spaceskip=\fontdimen2\font plus
\BIBentryALTinterwordstretchfactor\fontdimen3\font minus
  \fontdimen4\font\relax}
\providecommand{\BIBforeignlanguage}[2]{{%
\expandafter\ifx\csname l@#1\endcsname\relax
\typeout{** WARNING: IEEEtran.bst: No hyphenation pattern has been}%
\typeout{** loaded for the language `#1'. Using the pattern for}%
\typeout{** the default language instead.}%
\else
\language=\csname l@#1\endcsname
\fi
#2}}
\providecommand{\BIBdecl}{\relax}
\BIBdecl

\bibitem{Trust1}
Y.~Yu, K.~Li, W.~Zhou, and P.~Li, ``Trust mechanisms in wireless sensor
  networks: Attack analysis and countermeasures,'' \emph{Journal of Network and
  computer Applications}, vol.~35, no.~3, pp. 867--880, 2012.

\bibitem{general2}
C.~A. Kerrache, C.~T. Calafate, J.-C. Cano, N.~Lagraa, and P.~Manzoni, ``Trust
  management for vehicular networks: An adversary-oriented overview,''
  \emph{IEEE Access}, 2016.

\bibitem{r3}
D.~Liu, J.~Ni, H.~Li, X.~Lin, and X.~Shen, ``Efficient and privacy-preserving
  ad conversion for v2x-assisted proximity marketing,'' in \emph{Proc. IEEE
  15th Int. Conf. Mobile Ad Hoc and Sensor Systems}, 2018, pp. 10--18.

\bibitem{r4}
D.~Ulybyshev, A.~O. Alsalem, B.~Bhargava, S.~Savvides, G.~Mani, and L.~B.
  Othmane, ``Secure data communication in autonomous v2x systems,'' in
  \emph{Proc. IEEE Int. Congress Internet of Things (ICIOT)}, Jul. 2018, pp.
  156--163.

\bibitem{r5}
M.~A.~S. Junior, E.~L. Cominetti, H.~K. Patil, J.~Ricardini, L.~Ferraz, and
  M.~V. Silva, ``Privacy-preserving method for temporarily linking/revoking
  pseudonym certificates in {VANETs},'' in \emph{Proc. 12th IEEE Int. Conf. on
  Big Data Science And Engineering}, 2018, pp. 1322--1329.

\bibitem{r7}
C.~Xu, H.~Liu, P.~Li, and P.~Wang, ``A remote attestation security model based
  on privacy-preserving blockchain for v2x,'' \emph{IEEE Access}, vol.~6, pp.
  67\,809--67\,818, 2018.

\bibitem{vx1}
Y.~Yang, Z.~Wei, Y.~Zhang, H.~Lu, K.-K.~R. Choo, and H.~Cai, ``{V}2{X}
  security: A case study of anonymous authentication,'' \emph{Pervasive and
  Mobile Computing}, 2017.

\bibitem{vx3}
M.~Villarreal-Vasquez, B.~Bhargava, and P.~Angin, ``Adaptable safety and
  security in v2x systems,'' in \emph{Proc. IEEE Int. Congress Internet of
  Things (ICIOT)}, Jun. 2017, pp. 17--24.

\bibitem{vx4}
A.~Kiening, D.~Angermeier, H.~Seudie, T.~Stodart, and M.~Wolf, ``Trust
  assurance levels of cybercars in v2x communication,'' in \emph{Proceedings of
  the 2013 ACM Workshop on Security, Privacy \&\#38; Dependability for Cyber
  Vehicles}, New York, USA, 2013, pp. 49--60.

\bibitem{general5}
K.~J. Ahmed and M.~J. Lee, ``Secure, lte-based v2x service,'' \emph{IEEE
  Internet of Things Journal}, 2017.

\bibitem{r1}
H.~Jung, K.~Lim, D.~Shin, S.~Yoon, S.~Jin, S.~Jang, and J.~Kwak, ``Reliability
  verification procedure of secured v2x communication for autonomous
  cooperation driving,'' in \emph{Proc. Int. Conf. Information and
  Communication Technology Convergence}, Oct. 2018, pp. 1356--1360.

\bibitem{r2}
A.~Chattopadhyay, U.~Mitra, and E.~G. Ström, ``Secure estimation in v2x
  networks with injection and packet drop attacks,'' in \emph{Proc. 15th Int.
  Symp. Wireless Communication Systems (ISWCS)}, Aug. 2018, pp. 1--6.

\bibitem{J3}
A.~M. Shabut, K.~P. Dahal, S.~K. Bista, and I.~U. Awan, ``Recommendation based
  trust model with an effective defence scheme for {M}{A}{N}{E}{T}s,''
  \emph{IEEE Transactions on Mobile Computing}, vol.~14, no.~10, pp.
  2101--2115, 2015.

\end{thebibliography}

\end{document}